\documentclass[sigconf]{acmart}

\usepackage{xspace}
\usepackage{subcaption}
\usepackage{graphicx} 
\usepackage{tikz}
\usepackage{xcolor}
\usepackage{arydshln}
\usepackage{float}

\graphicspath{ {./images/} }

\AtBeginDocument{%
  }

\setcopyright{acmlicensed}
\copyrightyear{2026}
\acmYear{2026}
\acmDOI{XXXXXXX.XXXXXXX}
\acmConference[DSSIM'26]{Doctoral Summer School in Immersive Media}{June 26--06, 2026}{Porto, Portugal}
\acmISBN{978-1-4503-XXXX-X/2026/06}

\begin{document}

\title{A Serious Game for Training Civil Participation in Flood Crisis Management}

\author{Solmaz Goodarzi}
\affiliation{%
  \institution{University of Hamburg}
  \city{Hamburg}
  \country{Germany}}
\email{solmaz.goodarzi@uni-hamburg.de}
\authornote{All authors contributed equally to this research.}

\author{Arseni Ivanov}
\affiliation{%
  \institution{Lund University}
  \city{Lund}
  \country{Sweden}}
\email{arseni.ivanov@cs.lth.se}
\authornote{All authors contributed equally to this research.}

\author{David Maruscsák}
\affiliation{%
  \institution{Université Paris-Saclay}
  \city{Paris}
  \country{France}}
\email{david.maruscsak@universite-paris-saclay.fr}
\authornote{All authors contributed equally to this research.}

\author{Diana Quitéria Sousa}
\affiliation{%
  \institution{FEUP/INESC TEC}
  \city{Porto}
  \country{Portugal}}
\email{up722517@up.pt}
\authornote{All authors contributed equally to this research.}

\renewcommand{\shortauthors}{Ivanov et al.}

\begin{abstract}
Floods are among the most frequent and destructive natural disasters, requiring citizens to make critical decisions under rapidly changing conditions. Most serious games related to flooding emphasize awareness, preparedness, and education in flood management, but there is still a research gap on how to train citizens to act under pressure, uncertainty, and stress.
In this paper, we position Virtual Reality (VR) serious games as a promising approach to addressing this gap by providing experiential and immersive training opportunities for citizens.
We propose a conceptual VR serious game that enables participants to experience realistic flood scenarios, practice decision-making, rehearse appropriate response behaviors, and develop response-related skills in a safe and controlled environment. In addition, we suggest a research agenda and an evaluation framework focused on flood risk awareness, response capabilities, and stress-related outcomes. The framework thoughtfully integrates subjective, objective, and physiological measures to assess the effectiveness of immersive flood-response training.
Finally, we present design, methodological, ethical, and user-experience considerations in VR-based disaster training, and outline future research directions for the development and evaluation of immersive serious games for flood emergency response.
The present study aims to contribute to the growing intersection of serious games, extended reality, and disaster risk reduction by establishing citizen-centered response training as a significant yet under-researched field.
\end{abstract}



\begin{teaserfigure}
  \includegraphics[width=\textwidth]{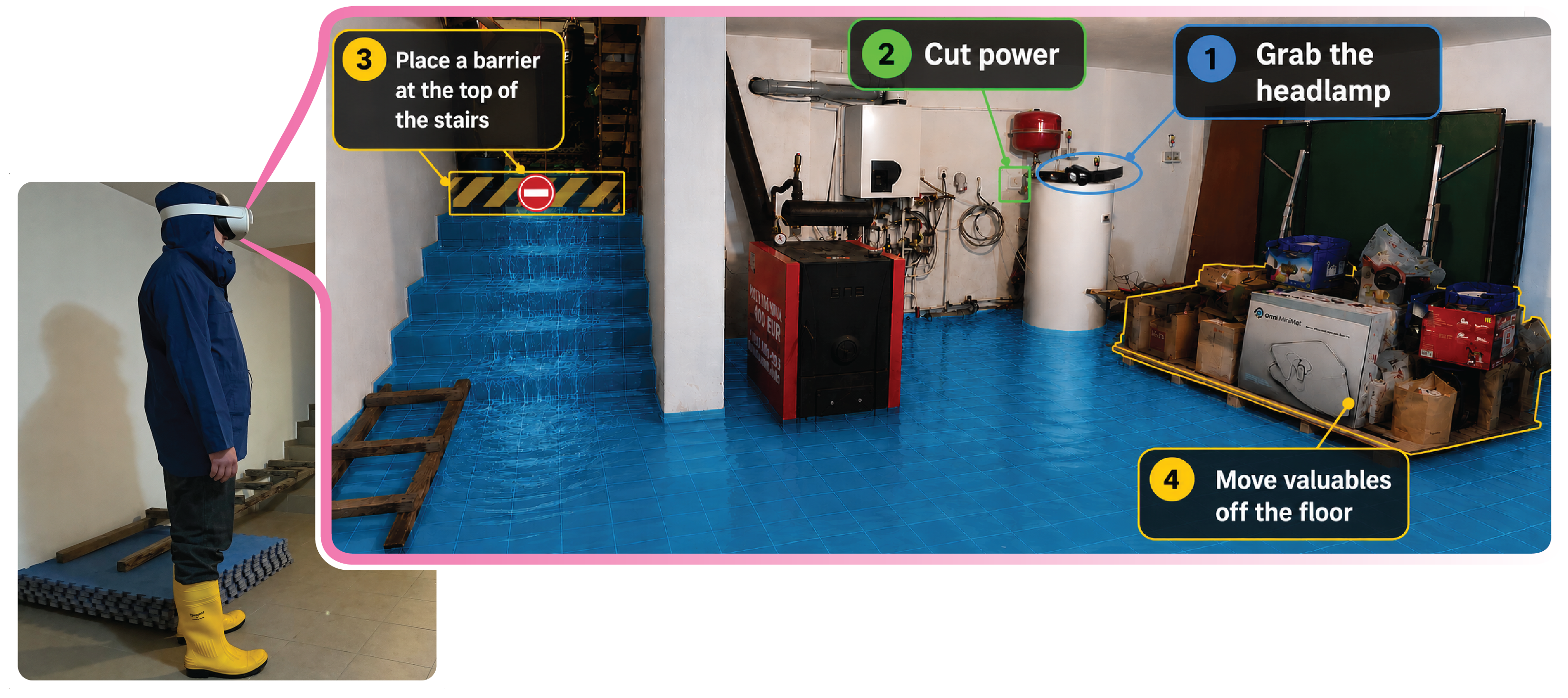}
  \caption{Our VR serious game places a participant (left) in a predefined training environment where they perform various response-phase tasks. In this scenario: (1) grab the headlamp, (2) cut the power, (3) place a barrier at the top of the stairs, and (4) move valuables off the floor.}
  \Description{Enjoying the baseball game from the third-base seats. Ichiro Suzuki preparing to bat.}
  \label{fig:teaser}
\end{teaserfigure}

\received{20 February 2007}
\received[revised]{12 March 2009}
\received[accepted]{5 June 2009}

\maketitle

\section{Introduction}

Floods are among the most frequent and destructive natural disasters worldwide, affecting millions of people every year and causing substantial social and economic damage \cite{intergovernmentalpanelonclimatechangeipccClimateChange20222023, debruijnFloodFatalityHazard2015, kvockaFloodHazardAssessment2016}. In recent years, climate change has further increased concerns about flooding, as rising sea levels and more frequent extreme rainfall events are expected to increase flood risks in many regions around the world \cite{intergovernmentalpanelonclimatechangeipccClimateChange20222023}. Recent flood events illustrate the scale of this challenge. For example, severe flooding across parts of Southeast Asia in 2025 resulted in more than 1,300 fatalities and affected millions of people, highlighting the devastating consequences such events can have on communities and societies \cite{apnews2025asiaflood}.

In these situations, emergency responders are not the only group that requires preparedness training. Citizens often have to make immediate decisions to protect themselves, their families, and, in some cases, to assist others around them. The actions taken during these critical moments can directly influence personal safety and the well-being of affected communities. 
Previous research suggests that prior flood-related experience is associated with higher levels of risk awareness and preparedness \cite{bradfordRiskPerceptionIssues2012, burninghamItllNeverHappen2008}, indicating that experiential aspects play an important role in how people perceive and respond to flood risks.
However, opportunities for citizens to gain practical experience with flood emergencies are inherently limited, as real flood events are unpredictable, dangerous, and difficult to recreate for training purposes.
Virtual Reality (VR) offers a promising way to address this challenge, allowing people to experience flood-related situations and explore response strategies without being exposed to real-world danger \cite{molVirtualFloodRisk2022, SIMPSON2022101764, 10.1145/3795011.3797385}.

Yet experiencing a VR scenario is not automatically the same as being prepared to act correctly once a flood is underway. 
Knowing that a flood is likely, or knowing what one is supposed to do in general, does not necessarily mean that a person can carry out the right sequence of actions in a chaotic, time-pressured, and often frightening situation. Research on disaster mitigation suggests that the effectiveness of people's actions depends not only on their disaster knowledge but also on the accumulation of practiced skills \cite{dpr.2023.26}. This distinction matters specifically for the response phase: while a leaflet or an awareness campaign can change what people know and how they perceive risk, it does not give them the chance to rehearse decision-making under the kind of stress, time pressure, and uncertainty that a real flood response entails. This is the gap we think VR serious games are particularly well suited to close, since they can recreate that pressure in a repeatable and safe way.

Existing flood-related serious games have primarily focused on increasing flood risk awareness, supporting preparedness, and improving understanding of flood management concepts \cite{10.1145/1581073.1581076, breuer2017exploring, DEMIRAY2025106418}, leaving comparatively less attention on citizen-centered training for the response phase, the period during which individuals must react to an unfolding flood, for example by responding to warnings, evacuating, seeking safety, or assisting others \cite{rodrigues2002dam}.
Building on these observations, we position VR serious games as a promising approach to prepare citizens for the response phase of flood emergencies. We argue that future flood-related serious games should move beyond awareness-oriented learning and preparedness-focused education by providing opportunities for citizens to actively experience unfolding flood situations, practice decision-making under pressure, and learn how to respond when they may need to protect themselves and their families, and potentially assist others. Such experiences may not only contribute to flood awareness and risk perception but also help individuals develop practical response-related skills and become more familiar with the cognitive and emotional demands associated with flood emergencies. 

To explore this position, we envision a VR serious game that places citizens in an unfolding flood emergency and allows them to experience situations that require observation, decision-making, and action (Figure~\ref{fig:teaser}). Through such an experience, participants can encounter realistic flood-related challenges, practice appropriate responses, and reflect on the consequences of their choices within a safe training environment. Based on this vision, we formulate the following overarching research question:

\begin{itemize}
\item \textbf{RQ:} How can VR serious games support the preparation of citizens for the response phase of flood emergencies?
\end{itemize}

To further investigate this question, we break it down into the following more specific research questions:

\begin{itemize}

\item \textbf{RQ1:} How does participation in a VR serious game influence citizens' flood risk awareness and perception?
\item \textbf{RQ2:} How does participation in a VR serious game influence citizens' ability to respond to flood emergencies?
\item \textbf{RQ3:} What is the impact of VR-based serious-game training for flood emergencies on participants' stress levels, as assessed by physiological, behavioral, and self-reported indicators?

\end{itemize}

Based on the literature on experiential learning, immersive training environments, and serious games, we formulate the following hypotheses:

\begin{itemize}

\item \textbf{H1:} Participation in a VR serious game will increase citizens' flood risk awareness and perception.
\item \textbf{H2:} Participation in a VR serious game will improve citizens' ability to respond to flood emergencies.
\item \textbf{H3:} VR-based serious-game training for flood emergencies leads to a significant decrease in stress responses from pre-test to post-test, as assessed through physiological, behavioral, and self-reported indicators.

\end{itemize}

Beyond these research questions and hypotheses, we envision several contributions to the fields of flood risk communication, serious games, and immersive training. First, we highlight citizen-centered response training as an underexplored direction within the flood serious game literature. Second, we propose the use of VR serious games as a means of preparing citizens not only to understand flood risks but also to actively respond to unfolding flood emergencies. Finally, we advocate for a broader evaluation perspective that considers awareness, response-related capabilities, and stress-related outcomes together when assessing flood emergency training interventions.

\section{Related Work}

We organize the surveyed literature along six dimensions and visualize where existing work concentrates and where our proposal sits (Figure ~\ref{fig:sankey}).

\subsection{Serious Games and Immersive Technologies for Training}
Serious games combine educational objectives with interactive gameplay, and they show consistent benefits for knowledge acquisition, content understanding, and learner engagement~\cite{boyle2016update, wouters2013meta}. Gamification, the addition of game elements to otherwise conventional material, improves learning relative to the same material without them~\cite{sailer2020gamification}. Immersive technologies extend these benefits. Virtual, augmented, and mixed reality training matches the effectiveness of traditional methods while also enabling training for dangerous, costly, or otherwise impractical situations~\cite{kaplan2021xr}. Therefore, virtual reality (VR) is well suited to evacuation and emergency-response training~\cite{feng2018ivr}.

\subsection{Games for Disaster Risk Reduction}
Serious games and simulations have proven useful for disaster risk reduction (DRR), particularly for raising awareness, identifying disasters, prompting preventive action, and helping players understand how a disaster affects other people~\cite{solinskanowak2018overview, gampell2016stopdisasters}. The online game Stop Disasters! lets players defend a settlement against disasters such as floods, earthquakes, and tsunamis by strengthening vulnerable houses and public buildings within a limited budget~\cite{gampell2016stopdisasters, solinskanowak2018overview}. Other titles train specific preparedness skills. Build a Kit has players assemble an emergency kit, and Supervolcano places them in emergency-response planning for a volcanic crisis~\cite{solinskanowak2018overview}. Many of these games are multiplayer, casting several participants as collaborating decision-makers rather than lone players experiencing the disaster in their own home environment~\cite{solinskanowak2018overview}.

\begin{figure}
    \includegraphics[width=\linewidth]{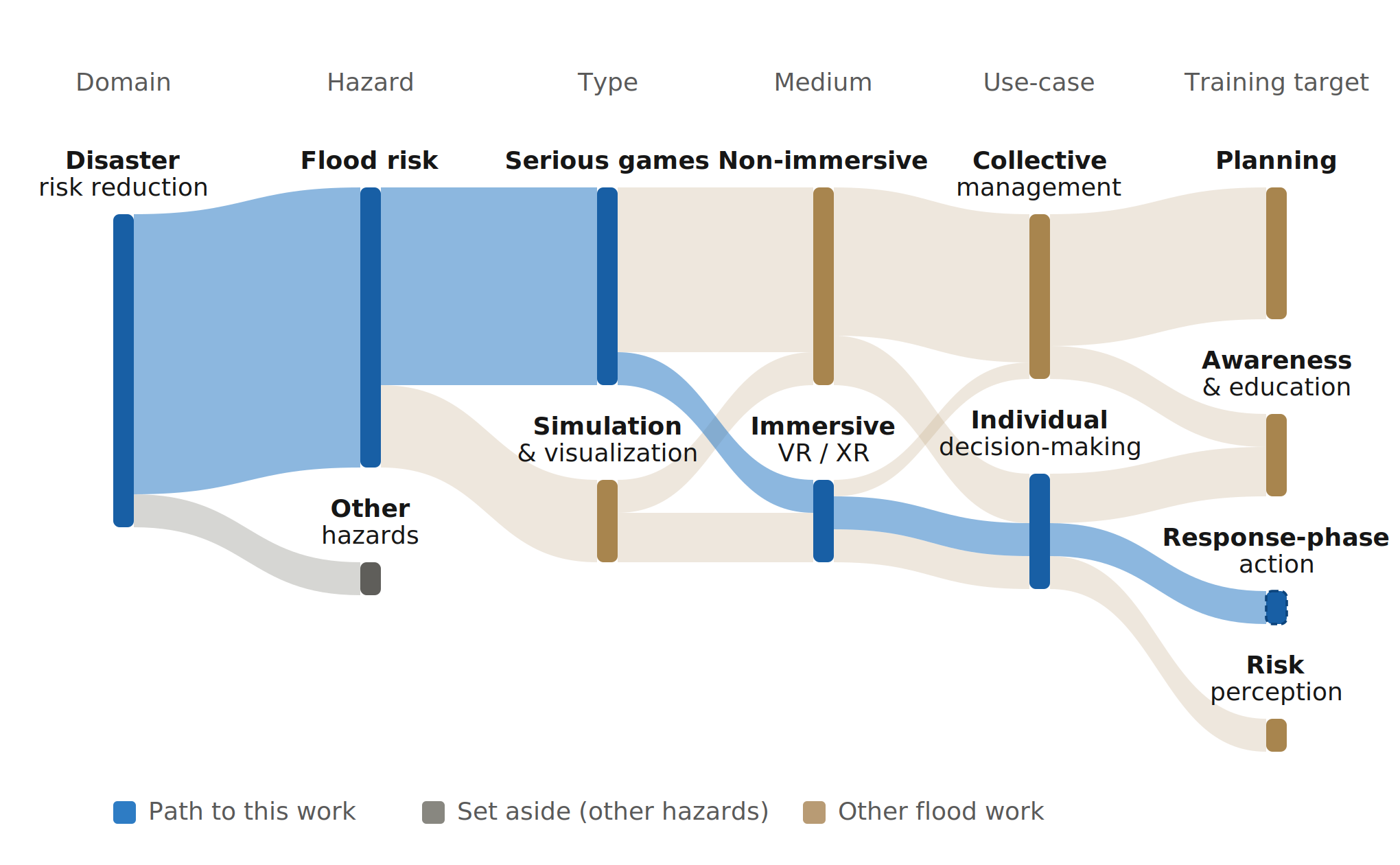}
    \caption{\textsf{An overview of the related work, mapping the surveyed literature across domain, hazard, type, medium, use-case, and training target. The blue path shows our proposal and highlights the gap it fills, targeting individual response-phase action left open by prior work.}}
    \label{fig:sankey}
\end{figure}

\subsection{Flood Management in Decision-Making Games}
A large share of flood-specific serious games focus on management and collective decision-making rather than individual action. Several designs cast players as planners who allocate budget, evaluate protective measures, and coordinate flood-risk strategy on the scale of a city~\cite{gamesreadytouse2018}. A related direction frames flood defence as a shared dilemma: the levee dilemma game studies how players manage protection under competing incentives~\cite{leveedilemma2023}, serious gaming has been applied to improve cooperation in flood-defence asset management~\cite{floodasset2022}, and game-theoretic models analyse how upstream and downstream actors share responsibility for prevention and financing~\cite{cooperativegame2019, gametheoretic2021}. Immersive systems extend this collective framing rather than the individual one, with extended-reality environments supporting shared situation awareness across control-room operators, civil-protection teams, and citizens during a flood event~\cite{xrsituationawareness2023}. Even citizen-facing designs in this tradition lean toward preparedness and coordination over in-the-moment action; the location-based game Flooded, for instance, builds flood preparedness by embedding awareness tasks in the player's own surroundings~\cite{flooded2014}. Across these works the player typically acts as a manager, planner, or coordinator deciding how to defend or proactively prepare a community, rather than as a citizen rehearsing what to do once a flood is already unfolding around them.

\subsection{Serious Games and VR for Flood Risk}
Flood risk has motivated many serious games for awareness, education, and planning, including awareness-raising games such as FloodSim~\cite{rebolledomendez2009floodsim}, collaborative training platforms~\cite{breuer2017secom}, games for exploring flood-mitigation measures~\cite{khoury2018serious}, interactive 3D environments~\cite{demiray2025floodgame}, and web-based games for younger audiences~\cite{mahamood2026floodedu}, which help players understand risk and plan interventions rather than rehearse how to act as a flood unfolds~\cite{forrest2022serious}. Beyond serious games, many VR systems simulate floods for risk communication without any game elements. Placing people inside a flood sharpens how accurately they perceive the danger and strengthens their intention to prepare~\cite{mol2022virtualflood, simpson2022stormsurge}, builds their confidence to evacuate~\cite{aksa2025immersive}, and reinforces safer behaviour through repeated practice~\cite{bettelli2023embankments}. Role-playing approaches further show that casting learners as different stakeholders in a flood scenario raises their awareness of the flooding's community-level impacts~\cite{spyropoulos2022narrative}. None of these approaches, however, trains the full sequence of decisions a citizen must make as a flood unfolds, from interpreting warnings and deciding under pressure to protecting their household and helping others. To address this gap, we propose a VR serious game that trains citizens for the response phase of a flood emergency (Figure ~\ref{fig:sankey}). Through a gamified experience, participants face an unfolding flood and practice decision making in a balanced setting that provides enough stress to keep them alert without overwhelming them, so the learning experience stays effective.

\section{Conceptual Approach and XR} 
The approach aims to position citizens as active agents in a flood scenario, interpreting environmental cues, assessing risks, and selecting actions in line with civil protection procedures. The flood crisis management is divided into three phases: preparedness (before), response (during), and recovery (after). Each phase contains decision points aligned with official guidelines and disaster behaviors, such as delayed evacuation, information seeking, and social influence. The decision points incorporate key behavioral patterns documented in disaster research, including delayed evacuation, information seeking, and social influence effects \cite{AitsiSelmi2015TheSF}. This structure guarantees procedural validity and facilitates the transfer of knowledge to real contexts.

\subsection{Theoretical Foundations}
The conceptual approach follows a serious games framework grounded in experiential learning and situated cognition to structure immersive VR training for flood preparedness, response, and recovery. Experiential learning, proposed by Kolb \cite{kolb1984}, involves four stages (concrete experience, observation and reflection, abstract conceptualization, and active experimentation) and refers to learning through direct experience in the context of learning. Awareness of variables such as social interaction and location enhances cognitive absorption and mental model change \cite{kwon2019,lui2022}. Situated cognition theory emphasizes that the learning context shapes cognitive outcomes \cite{lui2022}.  These frameworks support the design of training experiences that ground abstract knowledge in environmental conditions and behavioral patterns.

The design emphasizes active human-computer interaction (HCI), drawing on embodied cognition to create game interactions that mirror real scientific inquiry, increasing participants' engagement in the learning process, as the body mediates learning content \cite{10.3389/fpsyg.2023.1292110}. Participants observe, simulate, and dialogue within the virtual environment. They do not rely on narrative delivery. VR framing uses head-mounted displays to control the environment, reduce cybersickness, manage cognitive load, and preserve presence throughout training.

Cybersickness remains a barrier to VR adoption, causing discomfort and potentially disrupting immersion. Individual susceptibility
requires more attention as well as gender balance in group studies \cite{10.1007/s10055-022-00638-2}. Accessible design in this study is important to support users with visual, motor, and age-related impairments through inclusive interfaces, audio guidance, alternative input methods, and customizable interaction settings, following inclusive design principles \cite{ISO9241-210-2019}.

\subsection {User Scenarios and Decision Points}
Virtual scenarios show how people react to emergencies, the stress
of time pressure, and urgent decisions to increase safety \cite{bettelli2023embankments}. Imminent flood situations increase stress and require clear warnings
about stress and cybersickness, with the right to withdraw. Combining stress exposure training (SET) and Virtual Reality (VR) induces
stress. VR can induce stress and be well received. SET groups show
lower stress and higher competence, with repeated VR exposure
desensitizing future stress in a simulated environment, underscoring the value of controlled stress exposure while highlighting the
need for safeguard \cite{10.1186/s41077-024-00308-0}.

A three-scenario flood training encompasses preparedness (before), response (during), and recovery (after), each with decision points grounded in civil-protection protocols. The game mechanics employ time travel and feedback loops to elucidate cause-and-effect relationships, thereby fostering causal reasoning without imposing critical consequences \cite{10585429}.

The preparedness context presents a scenario in which the decision-maker must select the most effective method for acquiring information through emergency guidelines, determine whether to initiate an evacuation in advance, and establish a hierarchy of priorities, balancing the value of assets and the safety of family members. The results of this decision-making process are then compared to the corresponding risk states, providing a feedback loop that enables the decision-maker to assess their level of readiness. In this first moment, the feedback mechanisms are instrumental in elucidating the consequences of each decision: correct choices yield household readiness, while errors result in transparent risk states that clarify learning objectives without resorting to punitive language.

In the second scenario, participants need to complete the following tasks: first, navigate safe evacuation routes; second, reunite with separated family members; and third, respond to hazards with dynamic environmental feedback reflecting prior choices. In the second moment, the environmental feedback to the participants is dynamic, with alterations in water level, visibility, and available routes reflecting the trainee's prior choices and current decisions.

In the third scenario, which corresponds to recovery, the following topics are discussed: the implementation of a structural assessment, the process of water removal, the management of contamination, and the decision on whether to coordinate with relevant authorities or take independent action. Furthermore, the decision between a community-based approach and a household-centered approach must be deliberated.

\subsection{Pedagogical Game Mechanics}
Serious video games use game mechanics combined with psychological and pedagogical principles to create interactive environments that facilitate learning, skill development, and sustained engagement \cite{encyclopedia6040083}. XR-based flood crisis training blends serious games, experiential learning, and citizen-centered design to teach preparedness, response, and recovery with controlled immersion, while prioritizing usability, safety, and transfer to real-world actions.
The design system combines three layers: (1) Game Mechanics as Pedagogical Structure, defining goals, diverse interactions, authentic contexts, immediate and explanatory feedback, adaptive learning, and safety-by-design \cite{info:doi/10.2196/82505}; (2) Calibrated immersion to optimize presence and learning, adjusting visual fidelity to task demands; and (3) Cognitive load management with learning aids (feedback, hints, guidance) and paced segments (3–5 minutes) to balance action and reflection.

Game Mechanics as Pedagogical Structure presents the flood training principles, such as specific civil protection objectives (e.g., "evacuate safely within time constraints"), and the diversity of interactions between participants, including moving through space, manipulating objects, communicating with NPCs, and interpreting environmental signals. Moreover, the contextual authenticity is relevant to make scenarios reference real neighborhoods, official emergency procedures, and common behavioral responses to crises. XR increases presence, defined as the sense of "being there," which strengthens engagement and learning outcomes \cite{10.1098/rstb.2009.0138}. However, the design prioritizes action and reflection, procedural validity, and transfer to real contexts, with paced experiences to manage cognitive load and reduce overload. Co-creation with HCI experts, engineers, psychologists, and Virtual Environments designers informs open-access guidelines for safety, risk awareness, and quality of life near rivers \cite{10.3389/fpsyg.2021.674171}. These guidelines inform the design of citizen-centered flood training experiences.

Feedback loops allow participants to see the immediate consequences of their actions and determine whether those actions comply with civil protection guidelines, enabling them to then reflect on and review the sequence of their decisions.
Increasing levels of adaptive difficulty, combined with another principle that provides support for decision-making options during the first attempts, while in subsequent attempts, this support is removed and more complex activities are introduced.
Finally, the system's safety features prevent digital motion sickness by limiting movement options and offering alternative data entry methods. It should also be noted that failure should not result in penalties, as this would increase participants’ frustration and disinterest.

Accessibility features include reduced motion and alternative inputs, plus multimodal feedback (spatial audio, visuals, and interactive objects) to boost presence and learning, as well as enhance immersion \cite{11172647}. XR enhances situational awareness and decision-making by supporting clear learning objectives, user-centered interfaces, and adaptive difficulty. So, a co-creative approach is adopted for the design of XR digital content, with the objective of fostering facilitation and expanding interdisciplinary collaboration \cite{Schlemmer2025Stories}. The conceptual model uses XR as an experiential medium to train decision-making, behavioral rehearsal, and reflective learning in flood-crisis contexts. The system reliably regulates sensory intensity, pacing, and interaction complexity. This reduces cognitive overload and the risk of cybersickness. This aligns with human-centered design principles that emphasize usability, accessibility, and adaptation to user needs. It is essential to prevent adverse reactions. Designers should avoid triggering intense light or color bursts during immersive training. Epilepsy risks from visual stimuli vary by user, so opt for moderated proximity cues, stable patterns, calm backgrounds, and high-contrast but non-jarring color schemes. Tools like ColorADD and Adobe Color can help select accessible palettes, with careful consideration of contrast, motion sensitivity, and visual pacing to ensure a safe learning environment \cite{Sousa2022Adaptability}.

Regarding immersion's presence and engagement, it is important to note that immersion improves learning, engagement, motivation, self-efficacy, and emotions, with benefits for cognitive load, particularly for declarative knowledge \cite{bs15101322}. Therefore, it is important to adjust visual fidelity to the type of task, for example, high fidelity for declarative tasks (warning signs) and moderate fidelity for procedural tasks (timed route navigation).

Considering the Cognitive Load Management and learning aids, it should be noted that the Learning support features (such as feedback, suggestions, and guidance) in a self-regulated VR environment improve cognitive abilities, practical skills, self-regulation, and engagement \cite{SOUSA2022EDU}, especially among lower-performing participants \cite{https://doi.org/10.1111/bjet.13512}. The flood training program uses three adaptive features. So, it provides feedback after each decision, procedural suggestions when needed, and guidance on progress, including completion status and remaining objectives. Each scenario is divided into 3 to 5-minute segments, with pauses designed for reflection to balance action and learning.

\subsection{Research Design Methodology for Procedural Validity}
The system must be co-created by a multidisciplinary team, through collaborative sessions with civil protection officials, emergency responders, occupational psychologists, and VR experts, to ensure scenarios and decision points reflect authentic procedures and observed crisis behaviors. Each decision pathway aligns with published disaster research, mapping to real-world behavioral outcomes to validate the training’s procedural relevance.

\subsection{Accessibility and Human-Centered Design}
Accessibility in flood training centers on body-function–driven design and inclusive features \cite{10.1145/3663548.3688530}: reduced motion options to reduce cybersickness; hand-gesture or controlled-based input; multimodal feedback with audio and haptic cues to complement visual signage; adjustable colors (namely to colorblind persons), text size and contrast for those who have visual impairments; and simplified interactions to users with limited VR experience, enhancing presence and learning for diverse populations.

\subsection{Technical Architecture}
The serious game should be developed in Unity to run on commercial VR hardware (Meta Quest 3, HTC Vive), delivering moderate-fidelity 3D flood-prone environments that support situated learning while avoiding cognitive overload. Modular components enable regional customization, and telemetry records decision timing, sequences, errors, and reflections for real-time adaptation and instructor dashboards.

\section{Reflection on Feasibility and Evaluation}
\begin{figure*}
    \includegraphics[width=\textwidth]{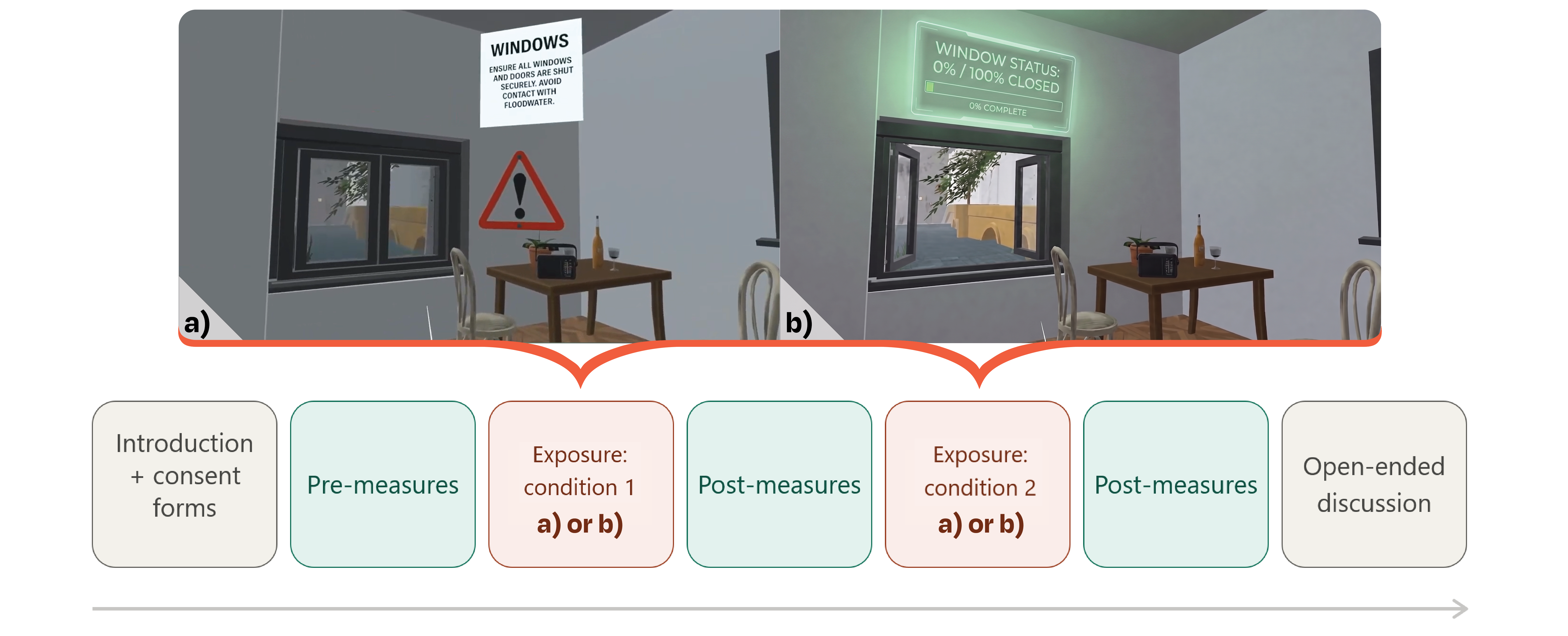}
    \caption{\textsf{The two training conditions (top) and the overall study procedure (bottom). Both conditions present the same flood task: (a) a conventional VR environment using static instructional signage, versus (b) a VR serious game with a gamified elements, such as progress feedback. In the procedure, participants complete pre-measures, both conditions in counterbalanced order (a or b) with post-measures after each, and a closing open-ended discussion.}}
    \label{fig:method}
\end{figure*}
To investigate the proposed position, we outline a preliminary research design to evaluate the effectiveness of VR serious games for citizen-centered flood response training.

\subsection{Study Design}
This study is designed to test the central position of this paper: that current flood serious games mainly cover awareness and preparedness, while the response phase has been largely left out. Our goal is to evaluate whether a VR serious game can fill that gap, and specifically whether gamification adds value beyond VR exposure alone.
We propose a mixed-methods empirical study using a within-subject design. This approach allows each participant to experience all training conditions while reducing the influence of individual differences. To minimize potential order effects, the presentation order of the conditions would be counterbalanced across participants.

The sample size will be informed by common practice in HCI research, where studies with a comparable within-subject design typically recruit around 12 participants per condition \cite{10.1145/2858036.2858498}. For our two-condition comparison, this suggests a target of approximately 20–25 participants, which is also consistent with related work evaluating VR-based flood and disaster training.
Participants will be recruited from the general public rather than domain experts or trained responders, since the goal of the study is to evaluate citizen-centered response training. No prior flood-related knowledge or experience will be required to take part, in line with the study's aim of preparing ordinary citizens, rather than already-trained individuals, for flood emergency response.

\subsection{Training Conditions}

To isolate the specific contribution of gamification, rather than VR exposure in general, the evaluation compares two VR-based training approaches that present identical flood-related content and learning objectives (Figure~\ref{fig:method}). The first condition would consist of a conventional VR training environment that provides instructional content and flood-related tasks without game design elements. The second condition would be a VR serious game that presents the same flood-related learning objectives while incorporating gamification elements such as challenges, feedback, rewards, and interactive decision-making mechanics.

This comparison directly addresses RQ2 and RQ3 by testing the hypothesis that game design elements improve citizens' response capability and influence participants' stress levels, as assessed through physiological, behavioral, and self-reported indicators, to a greater extent than conventional VR-based flood training.

The two conditions would both focus on the issue of citizen participation during the response phase of a flood emergency. Participants would encounter situations requiring observation, decision-making, self-protection, and helping behaviors under evolving emergency conditions. Each scenario would require participants to identify hazards, select appropriate protective actions, and respond within realistic time constraints. These scenarios would provide the behavioral evidence required to evaluate response capability (RQ2) while creating the conditions needed to assess stress responses through physiological, behavioral, and self-reported indicators (RQ3).

\subsection{Procedure}
Participants would first receive an introduction to the study and provide informed consent (Figure~\ref{fig:method}). Pre-measures would then be collected to establish baseline values for the investigated constructs.

After the pre-assessment, participants would complete both training conditions in counterbalanced order. Following each condition, post-measures would be administered. At the end of the study, participants would take part in an open-ended interview to provide qualitative feedback regarding their experiences, perceived realism, learning outcomes, and overall impressions of the training approach.


\subsection{Measures}
To evaluate the proposed research questions and hypotheses, multiple complementary measures would be collected, each selected to provide evidence relevant to a specific research question.

To answer RQ1 (flood risk awareness and perception), the study would collect subjective measures of flood risk awareness and perception before and after each training condition.
To answer RQ2 (response capability), the study would combine objective performance measures, task completion time, response accuracy, success rate, and decision quality during flood-related scenarios, with participants' subjective ratings of workload and sense of presence, since both contribute to understanding how well participants were able to respond within the simulated environment.
To answer RQ3, which concerns the impact of VR-based serious-game training on participants' stress levels, the present study proposes a methodological approach that will combine self-reported, behavioral, and physiological indicators of stress. Self-reported stress would be assessed using validated questionnaires, behavioral indicators would include response latency, hesitation during decision-making, and task completion time, and physiological measures would include heart rate (HR) and galvanic skin response (GSR). The implementation of these complementary measures would facilitate the capture of both perceived and objectively observed stress during flood emergency training.
Game experience would additionally be assessed as a subjective measure to characterize participants' engagement with the gamified condition specifically.
Combining subjective, objective, and physiological measures across these research questions would allow a comprehensive evaluation of both learning outcomes and user experience.

\subsection{Data Analysis and Expected Outcomes}
The collected data could be analyzed using a mixed-methods approach that combines quantitative and qualitative evidence. Quantitative analyses would investigate differences in awareness, response capability, and stress-related outcomes between conditions, while qualitative feedback would provide additional insights into participants' experiences and perceptions.

Such an evaluation would help assess whether VR serious games can provide benefits beyond conventional VR training and contribute to the preparation of citizens for flood emergency response situations.

\section{Ethical, UX and Assessment Considerations}
The VR serious game aims to provide a realistic, safe, and feasible flood-emergency training experience for citizens facing flood emergencies, balancing immersion, usability, safety, and research needs throughout design and evaluation. This section outlines the key user-experience design, followed by ethical considerations and a methodology aligned with the research questions.

\subsection{User Experience and Design Considerations}
The proposed training consists of two VR conditions that present identical learning objectives through different interaction approaches. One traditional VR training and the other with a serious game version, each with short flood scenarios that gradually increase decision complexity while minimizing fatigue and cybersickness. 

Each scenario places participants in situations that require observation, risk assessment, decision making, self-protection, and, when appropriate, helping other citizens. Success depends on selecting the right actions within realistic time constraints, as outlined in official flood emergency guidelines. Performance indicators clearly include decision accuracy, successful completion of critical safety actions, response time, and overall scenario completion.

The interface follows human-centered design principles, prioritizing clarity, accessibility, and low cognitive load. Reduced-motion navigation, alternative interaction methods, multimodal feedback, and configurable interaction settings accommodate participants with different levels of VR experience and physical abilities. The design is clear: it regulates sensory intensity and pacing to maintain immersion without overwhelming participants or increasing the risk of cybersickness \cite{ISO9241-210-2019,computers11040051,10.1007/s10055-022-00638-2}.

\subsection{Informed Consent and Data Privacy}
The study requires that participants provide informed consent before participating. Participants are informed about the objectives of the study, the risks involved, the stress exposure, and the possibility of cybersickness. Participants may withdraw from the study at any time without incurring any penalties.

The study excludes pregnant individuals, participants undergoing treatment for flood-related trauma or panic disorders, individuals with severe psychological conditions, severe cardiovascular diseases, epilepsy, and those known to be susceptible to cybersickness. It is evident that these criteria can reduce health risks during immersive flood simulations. These exclusion criteria serve to protect participants during the preliminary evaluation. It is recommended that subsequent research endeavors encompass a broader range of participants by adapting the experience. Lower-intensity scenarios, alternative interaction techniques, desktop versions, or augmented reality applications could support the inclusion of children, older adults, and individuals with specific health conditions while maintaining participant safety.

The system employs anonymous participant identifiers and encrypted repositories to store all collected data. The study exclusively collects the information necessary to evaluate the proposed research questions, in accordance with data minimization principles. Physiological measurements are subject to protection measures, because biometric signals may enable participant re-identification when combined with other datasets \cite{Sawant2023BiometricPrivacy}.

\subsection{Performance Assessment}
The system collects objective performance metrics, including task completion time, response speed, decision accuracy, and success rate, to assess learning outcomes. Task completion time captures overall efficiency, while response speed measures reaction latency to specific prompts. These complementary metrics enable performance assessment.

\subsection{Evaluation Instruments}
The evaluation methodology addresses the research questions introduced in Section 1. Each research question calls for a distinct set of evidence, needing the integration of subjective, objective, and physiological measures to evaluate the proposed training approach.

The primary research question (RQ1) investigates whether participation in the virtual reality (VR) serious game enhances flood risk awareness and perception. The study draws parallels between pre- and post-training questionnaires, using them to assess participants' comprehension of flood risks, perceived preparedness, and confidence in responding to flood emergencies.

RQ2 explores citizens' capacity to respond to flood emergencies. The evaluation process involves analyzing objective performance indicators collected during the training scenarios. These indicators encompass factors such as decision accuracy, response time, successful completion of critical safety actions, and completion of the scenario. The present study uses the NASA Task Load Index (NASA-TLX) and the Igroup Presence Questionnaire (IPQ) to facilitate a more comprehensive interpretation of participants' performance in the immersive environment \cite{10.1177/29941520251415449}.

RQ3 evaluates the impact of the VR serious game on participants' stress levels through physiological, behavioral, and self-reported indicators. The assessment of self-reported stress is conducted using the Short Stress State Questionnaire (SSSQ). Physiological measures encompass galvanic skin response (GSR), heart rate (HR), heart rate variability (HRV), pupil dilation, electroencephalography (EEG), electrodermal activity (EDA), respiration (RESP), and body temperature. Behavioral indicators include response latency, hesitation during decision-making, and task completion time. The Simulator Sickness Questionnaire (SSQ) is a tool designed to assess the distinction between stress responses and discomfort experienced during immersive virtual reality.  A novel weighted aggregation approach integrating refined metrics for EDA, HRV and temperature-related stress factors offers a comprehensive stress evaluation \cite{11172566}. After acquiring physiological data from the sensors, it is processed to extract time, frequency, and nonlinear features, as well as GSR and RESP features. The research results show that HR approximate entropy, the low-to-high frequency ratio, low frequency, standard deviation, GSR, and RESP are reduced, and high frequency and improved with stress management interventions \cite{bioengineering10070766}.

Collectively, these complementary measures help to answer each research question while evaluating the feasibility and potential effectiveness of citizen-centered flood response training through VR serious games. However, careful interpretation is required due to sensor noise and the context in which the experience occurs, which can affect the analysis and corresponding evaluations (baseline arousal level, VR experience, temperature, movement). However, it is possible to evaluate all data to extract information on usability, engagement, workload, stress, and learning effectiveness.

\section{Future Research Plan}
The study described in the previous section is the first step of a longer research effort. Before it can take place, the prototype needs to go through further design iterations. We plan to keep the design cycle used so far, in which the scenario content is based on official emergency guidelines and each version of the game is reviewed with domain experts and improved based on their feedback. We also want to involve citizens in this process through co-creation sessions, so that the scenarios reflect how people actually think and behave during a flood, and not only how official procedures describe it. Pilot tests will be used to adjust the pacing and sensory intensity of the experience, so that the stress we measure later comes from the simulated emergency and not from discomfort with the medium \cite{computers11040051}. 

The within-participant study can only capture effects measured directly after training. Since flood events are rare, the value of citizen training depends on whether the awareness, response skills, and stress regulation it builds last until they are actually needed in practice. We therefore plan to extend the evaluation into a longitudinal study, in which the post-measures are reported weeks and months after the training session, addressing a common weakness of VR-training research that typically captures only immediate post-test effects. In the same follow-up phase, the serious game should be compared with traditional training formats such as informational material and physical drills. Traditional training is often hard to motivate and fund at the scale of the general public, so showing what a serious game adds, both in learning outcomes and in engagement, would support its use as a more accessible alternative.

Another question the planned study leaves open is how individual design choices contribute to the measured outcomes. The comparison between the gamified and non-gamified condition treats gamification as a whole. Follow-up studies can change single mechanics, such as time pressure, feedback, and rewards, to study how each of them affects engagement, learning, and stress \cite{10.1145/950566.950595, bioengineering10070766}. The results will also need to be confirmed across different regions, since flood risks and prior exposure differ between places.

\section{Discussion and Future Directions}

We have argued that VR serious games are a promising and underexplored way to prepare citizens for the response phase of flood emergencies, and we have outlined a study design to test this position. The approach has clear limitations. A VR experience is not reality and cannot cover every possible flood scenario, and it is hard to validate whether trained behavior carries over to real emergencies, since training effectiveness cannot ethically be tested in an actual crisis. The training is also an individual experience, while real flood crisis management depends on cooperation between citizens, responders, and authorities. Gamification itself carries a risk: the same elements that make training engaging may weaken the stress-driven survival instincts that guide behavior in a real emergency. Prior work shows VR can improve objective risk understanding, such as estimating water levels, while at the same time inducing riskier behavioral intentions such as delaying evacuation \cite{SIMPSON2022101764}. This tension is one of the reasons why the proposed evaluation measures stress alongside awareness and performance, and why it compares the game with a non-gamified VR condition.

The exclusion criteria needed for safe experimentation leave out groups that are among the most vulnerable in a flood, including children, older adults, and persons with disabilities. Reaching them requires dedicated, more inclusive designs with accessible interfaces, audio guidance, and alternative input methods \cite{ISO9241-210-2019}, rather than one experience adapted after the fact. Training needs also differ with expertise, and future work should consider how the experience should differ between laypersons, community volunteers, and professional responders. We believe that targeting a different group, such as children, will require specific and non-generalizeable settings that will require an all-together different research proposal. 

On the technical side, Augmented Reality \cite{10.1162/pres.1997.6.4.355} would allow the training to be personalized to the participant's own environment, letting citizens rehearse flood responses in their own homes and neighborhoods. Such personalization raises a privacy question of its own, as people may be sensitive about sharing data about their houses and private spaces, so the system should only collect as much information as the training actually needs. Cooperative multiplayer experiences are another natural extension, since a single-player game cannot capture the collaborative side of a real flood response that might be of interest when we consider helping others in such situations. They would also open new design questions, for example how players should weigh their own safety against the needs of the community when deciding whom to help first. 

We hope that this position encourages the serious games, XR, and disaster management communities to move beyond awareness-focused flood education and toward citizen-centered response training that prepares people to take action in the moment of the disaster.

\begin{acks}
The authors received funding from the European University Alliance for Global Health (EUGLOH)
This paper's writing process was supported by OpenAI's GPT for improving clarity
\end{acks}

\bibliographystyle{ACM-Reference-Format}
\bibliography{sample-base}

\end{document}